\documentstyle[sprocl]{article}

\def\Journal#1#2#3#4{{#1} {\bf #2}, #3 (#4)}

\def\PRD{{\em Phys. Rev.} D}

\def\be{\begin{equation}}
\def\ee{\end{equation}}
\def\bea{\begin{eqnarray}}
\def\eea{\end{eqnarray}}

\begin{document}

\title{GRAVITY AND  INSTANTONS  }

\author{ M. YU. KUCHIEV  }

\address{
School of Physics, University of New South Wales,
Sydney, 2052, Australia,\\
Email: kmy@newt.phys.unsw.edu.au}


\maketitle
\abstracts{Conventional non-Abelian SO(4)
gauge theory is  able 
to describe gravity provided the gauge field 
possesses a specific polarized vacuum
state in which the instantons have a
preferred orientation. Their orientation
plays the role of the order parameter for the 
polarized phase of the  gauge field.
The interaction of a  weak and smooth gauge field
with the polarized vacuum is described by an effective 
long-range action which is identical to the 
Hilbert action of general relativity.
In the classical limit this action results in 
the Einstein equations of general relativity.
Gravitons appear as the 
mode describing propagation of the gauge field which strongly 
interacts with the oriented  instantons.
The Newton gravitational constant describes the
density of the considered phase of the gauge field. The 
radius of the instantons under consideration is comparable with the 
Planck radius. }

\section{Instantons in an external field}
\label{int}

This work reviews the idea proposed in \cite{IAP,det}
which  suggests a new explanation for the origin of gravitational forces.
In the proposed scenario gravity arises as a particular effect 
in  the conventional 
Yang-Mills gauge theory \cite{YM}
formulated in flat space-time.
All geometrical objects 
necessary for the gravitational phenomenon
can originate from the gauge degrees of freedom
if a particular nontrivial vacuum
state, which we will call the vacuum with polarized instantons
develops in the $SO(4)$ gauge theory.

Consider Euclidean formulation of the  $SO(4)$ gauge theory.
The gauge algebra for $SO(4)$ gauge group consists of  two $su(2)$ 
gauge subalgebras, $so(4) = su(2)+su(2)$. 
The  instantons and antiinstantons  can belong to any one
of these two available $su(2)$ gauge subalgebras.
It is convenient to  choose the generators for one 
$su(2)$ gauge subalgebra to be $(- 1/2)\eta^{aij}$ and the 
generators for the other  one
to be $(-1/2) \bar \eta^{aij}$. 
To distinguish between these two subalgebras we will  refer to  them as 
$su(2)\eta$ and $su(2) \bar \eta$. 
Symbols $\eta^{aij}, \bar \eta^{aij}$
are the usual 't Hooft symbols, $a =1,2,3,~ i,j = 1, \cdots,4$.
In this notation
the gauge potential and the strength of the gauge field are
\begin{eqnarray} \label{Af}
A^{ij}_{\mu}&=&-\frac{1}{2} (A^a_{\mu} \eta^{aij} + 
\bar A^a_{\mu} \bar \eta^{aij}), \\ \label{F}
F^{ij}_{\mu \nu}&=&-\frac{1}{2} (F^a_{\mu \nu} \eta^{aij} + 
\bar F^a_{\mu \nu} \bar \eta^{aij}),
\end{eqnarray}
where $A^a_{\mu}$ and $F^a_{\mu \nu}$ belong to $su(2) \eta$ and 
$\bar A^a_\mu,\bar F^a_{\mu\nu}$ belong to $su(2) \bar \eta $.
The Yang-Mills action reads
$
S= 1/(4g^2)\int F^{ij}_{\mu\nu}(x)F^{ij}_{\mu\nu}(x) \, d^4x.
$
The Latin indexes $i,j=1, \cdots,4$ label the 
isotopic indexes, while the Greek indexes $\mu,\nu=1,\cdots,4$
label the indexes in Euclidean coordinate space.
Remember that we consider the usual 
gauge field theory in  flat space-time.
For the chosen normalization of generators  
the relation between the gauge potential and the field strength reads
$ F_{\mu\nu}^{ij}=\partial_\mu A_{\nu}^{ij}-
\partial_\nu A_{\mu}^{ij}
+A_{\mu}^{ik}A_{\nu}^{kj}- A_{\nu}^{ik}A_{\mu}^{kj}.
$

Consider interaction of an  instanton
with an external gauge field which has trivial topological structure
and is weak and smooth.
Thus formulated problem was  first considered 
by Callan, Dashen and Gross \cite{CDG} where
it was shown  that the interaction of an instanton 
with the external field is described by an effective action
\begin{equation}\label{cdg}
\Delta S = 
\frac{2 \pi^2 \rho^2}{g^2}\bar \eta^{a\mu\nu} D^{ab}F^b_{\mu\nu}(x_0).
\end{equation}
Here $\rho$ and $x_0$ are the radius and position of the instanton.
The matrix  $D^{ab} \in SO(3)$ describes the orientation of the instanton 
in the $su(2)$ gauge subalgebra where the instanton 
belongs.
$F^b_{\mu\nu}(x)$
is the gauge field in the subalgebra where the instanton belongs.
This field has to be taken in the singular gauge \cite{tH}.
The  interaction of an antiinstanton with an external field
is described similarly. The only distinction
is that it produces the corresponding term with the 't Hooft symbol 
$\eta^{a\mu\nu}$ instead of $\bar\eta^{a\mu\nu}$  which stands in
(\ref{cdg}). 

Let us now generalize the  problem. Suppose that there
is a number of instantons and antiinstantons which
belong to either $su(2)\eta$ or $su(2)\bar\eta$ gauge
algebras. 
Assuming that the dilute gas approximation is valid
we  find that the contribution to the action reads
\begin{equation} \label{4gas}
\Delta S = -\frac{\pi^2}{g^2} \sum_k
\eta^{A\mu\nu}\eta^{Bij}  
T^{AB}_k \rho^2_k F^{ij}_{\mu\nu}(x_k).
\end{equation}
Here  $k$ runs over all instantons and antiinstantons
which have radiuses and coordinates $\rho_k$ and $x_k$.
To simplify notation
the 't Hooft symbols are enumerated 
as 6-vectors $\eta^A = (\eta^a,\bar \eta^b),
~A=1,\cdots,6;~
a,b=1,2,3$.
To describe an orientation of every (anti)instanton
it is convenient to introduce a $6\times 6$ matrix $T^{AB}_k,~~
A,B=1,\cdots,6$ 
\begin{equation} \label{CAB}
T_k^{AB} \equiv T_k =
\left(  \begin{array}{cc} C_k &  D_k \\ 
 \bar D_k &   \bar C_k \end{array} \right)
\end{equation}
as a set of four $3\times 3$ matrixes  $C_k,\bar C_k,D_k,\bar D_k$.
For any given $k$-th (anti)instanton only one of these four matrixes 
is essential while the other three are
equal to zero.  This nonzero matrix  belongs to $SO(3)$ and  describes the 
orientation of the $k$-th  topological object in the gauge algebra where
it belongs.
For example, if the $k$-th object is 
an  antiinstanton  in the  $su(2)\eta$ gauge subalgebra, 
then we assume that $C_k\in SO(3)$ describes its orientation 
in the  $su(2)\eta$ while $\bar C_k,D_k,\bar D_k =0$.

Let us consider the behavior of the
ensemble of instantons in the vacuum state assuming that
there exists  a weak and smooth topologically trivial gauge field
$F^{ij}_{\mu\nu}(x)$. One can derive the
effective action which describes 
interaction of the vacuum with this field
averaging (\ref{4gas}) over
short-range quantum fluctuations in the vacuum.
The result can be presented as  the effective action
\begin{equation} \label{f/4}
\Delta S = 
-\int \eta^{A\mu\nu}\eta^{Aij}{\cal M}^{AB}(x)F^{ij}_{\mu\nu}(x)
\,d^4 x.
\end{equation}
where the matrix ${\cal M}^{AB}(x)$ is
\begin{equation} \label{M^AB}
{\cal M}^{AB}(x) = 
\pi^2 \langle \,
\frac{1}{g^2} 
\rho^2  T^{AB}  n(\rho, T,x)\,\rangle.
\end{equation}
The brackets $\langle\, \rangle$ here describe 
averaging over quantum fluctuations whose
wavelength is shorter than a typical distance
describing variation of the external field.
These fluctuations in the dilute gas approximation for instantons 
should include averaging over positions,
radiuses and orientations of instantons. 
In (\ref{M^AB}) $ n(\rho, T,x)$ is the concentration
of (anti)instantons which have 
the radius $\rho$ and the orientation described
by the matrix $T\equiv T^{AB}$. In the usual vacuum 
states the concentration of instantons does not depend on
the  orientation, $n(\rho, T,x) \equiv n(\rho, x)$. In that case
an averaging over orientations gives the trivial result
${\cal M}^{AB}(x) \equiv 0$, as mentioned by
Vainshtein {\it et al}.~\cite{VZNS} 
The  main goal of this paper is to investigate
what happens if the concentration $n(\rho, T,x) $ does depend on the 
orientation $T=T^{AB}$ providing the nonzero value for the matrix 
${\cal M}^{AB}(x)$.

It is shown  below that interesting physical
consequences  arise if one assumes that
the matrix ${\cal M}^{AB}(x)$ satisfies
the following  condition 
\begin{equation} \label{M=fM}
{\cal M}^{AB}(x)  = \frac{1}{4}\,f\, M^{AB}(x),
\end{equation}
where $f$ is a positive constant and
$M^{AB}(x) \in SO(3,3)$, which means that 
$ M \Sigma M^T= \Sigma$, where
 the matrix $\Sigma^{AB}$ is defined as
\begin{equation}\label{sigma}
\Sigma  =
\left(  \begin{array}{cr} \hat 1 &  \hat 0 \\ \hat 0 & -\hat 1
\end{array} \right).
\end{equation}
Condition  (\ref{M=fM}) is  the main {\it assumption} 
about   properties of
the vacuum state of the $SO(4)$ gauge theory.
Now we can  clarify the meaning of
the term  ``polarization of instantons'' which was introduced
above intuitively.
We  say that there is the polarization of instantons
with  the $SO(3,3)$ symmetry,  if
(\ref{M=fM}) is  valid. 
One can interpret  (\ref{M=fM})
as the statement that there exists the new nontrivial
phase of the $SO(4)$ gauge theory. The matrix
$M(x)$ plays the role of the order parameter for this phase.
The (anti)instantons which contribute to the 
nontrivial value of the matrix $T^{AB}$ in (\ref{M^AB}) can
be looked at as a specific condensate.
The constant $f$ characterizes the density of this condensate.

Notice that existence of a state with polarized instantons 
does not  come into contradiction  
with general principle of gauge invariance
which in  the context considered is 
known as the Elitzur theorem,~ \cite{eli} see also  
the book of Polyakov \cite{pol} and recent lectures of 
Hamer ,~\cite{ham} because
the orientation of an instanton is a gauge invariant
parameter.

It is 
instructive to re-write (\ref{f/4}) in another form.
Notice  that there exists
a relation between matrixes belonging 
to $SO(3,3)$ and matrixes
belonging to $SL(4)$ groups which is well known, see
a book of Gilmore \cite{gil}.
It can be presented as a statement that 
for any $M^{AB} \in SO_+(3,3),~A,B=1,\cdots,6$
there exists some matrix $H^{i\mu}\in SL(4),~i,\mu=1,\cdots,4$ satisfying
\begin{equation} \label{hom}
H^{i\mu}H^{j\nu}-H^{i\nu}H^{j\mu} =
\frac{1}{2} \eta^{A\mu\nu}\eta^{Bij}{M}^{AB}.
\end{equation}

Identifying $M^{AB}(x)=M^{AB}$ one finds from (\ref{hom}) 
$H^{i\mu}=H^{i\mu}(x)\in SL(4)$ which 
can be considered as another representation for the
order parameter.
Substituting  $H^{i\mu}(x)$ defined by
 (\ref{hom}) into (\ref{M=fM}) one can rewrite
the action (\ref{f/4})  
in the following  useful form
\begin{equation} \label{HHF}
\Delta S = -f \int H^{i\mu}(x)H^{j\nu}(x)F^{ij}_{\mu\nu}(x)
\,d^4x.
\end{equation}
Up to now our consideration was restricted by
orthogonal coordinates.
It is instructive however to present the action 
(\ref{HHF}) in arbitrary coordinates. 
Under the coordinate transformation
the matrix $H^{i\mu}(x)$ is transformed as
$
H^{i\mu}(x)\rightarrow h^{i\mu}(x') = 
{ (\partial x'^\mu}/{\partial x^\lambda)}  H^{i\lambda}(x)
$, where
the transformed matrix is called
$h^{i\mu}(x)$. Using it one can
present the action  (\ref{HHF}) 
in arbitrary coordinates $x$ 
in the following
form
\begin{equation} \label{hhF}
\Delta S = -f \int h^{i\mu}(x)h^{j\nu}(x)F^{ij}_{\mu\nu}(x) \det h(x)
\,d^4x.
\end{equation}
We use notation
in which $h^{i}_\mu(x)$ is understood
as the matrix  
inverse to $h^{i\mu}(x)$,  i.e.
$
h^{i}_\mu(x)h^{j\mu}(x) = \delta_{ij}
$.
The determinant in (\ref{hhF}) is defined as a determinant of
this inverse matrix, $\det h = \det [h^i_\mu]$.
Thus the factor $\det h(x)$ in  (\ref{hhF})
simply accounts for the variation of the
phase volume under the coordinate transformation.

\section{The Riemann geometry and the Einstein equations}
\label{eineq}
 Excitations above the polarized vacuum 
should possess interesting properties 
because  variation of the gauge field results in the contribution
to the action (\ref{hhF}) 
which is linear in the  field. This is in contrast to
the standard quadratic behavior of the conventional Yang-Mills action.
Let us examine the properties of excitations in the classical
approximation.

We will assume that the fields considered
vary on the macroscopic distances, say $\sim 1$cm, and 
their magnitude can be roughly estimated as 
$|F^{ij}_{\mu\nu}| \sim 1/ {\rm cm^2}$.
We will see below that the constant $f$ which was
defined in (\ref{M^AB}),(\ref{M=fM}) is  large,
$f \sim 1/r_{\rm P}^2$, where $r_{\rm P}$ is the Planck radius.
This shows that for weak fields the integrand in the 
Yang-Mills action   is suppressed compared to 
the Yang-Mills action by a drastic factor 
$ |F^{ij}_{\mu\nu}|/f \sim (r_{\rm P}/1 {\rm cm})^2 = 10^{-64}$.
Therefore
our  priority is to take into account
the action (\ref{hhF}) which describes interaction
of the weak field with polarized instantons, neglecting the Yang-Mills
action. 

Let us consider
the action (\ref{hhF}) as  a functional which depends on the weak field vector
potential and the matrix $h^{i\mu}(x)$ which describes orientations
of instantons,
$ \Delta S =\Delta S( \{A^{ij}_\mu(x)\},\{ h^{i\mu}(x)\})$.
Classical equations for this functional  read
\begin{eqnarray} \label{dsda} 
\frac{\delta( \Delta S)}{ \delta A^{ij}_{\mu}(x)} &=& 0,
\\ \label{dsdh}
\frac{\delta( \Delta S)}{ \delta h^{i \mu}(x)} &=& 0.
\end{eqnarray}
In order to present classical equations
in a  convenient form let us define three
quantities, $g_{\mu \nu}(x), 
\Gamma^{\lambda}_{\mu \nu}(x)$, and 
$R^{\lambda}_{\rho \mu \nu}(x)$:
\begin{eqnarray} \label{ghh}
g_{\mu \nu}(x) &=& h^{i}_{\mu}(x) h^{i}_{\nu}(x),
\\ \label{gam}
 \Gamma^{\lambda}_{\mu \nu}(x) &=& 
h^{i \lambda}(x)h^{j}_{\mu}(x)A^{ij}_
{\nu}(x) + h^{i \lambda}(x) \partial_{\nu} h^{i}_{\mu}(x),
\\ \label{RF}
R^{\lambda}_{\rho \mu \nu}(x) &=& h^{i \lambda}(x) 
h^{j}_{\rho}(x)F^{ij}_
{\mu \nu}(x).
\end{eqnarray}
After simple calculations
the first classical 
equation (\ref{dsda}) may be presented in the  form
$ \Gamma_{\lambda\mu}^{\sigma}(x)=1/2 
g^{\sigma\tau}(x) \left[
\partial_\lambda g_{\tau \mu }(x)+
\partial_\mu g_{\lambda \tau }(x)-
\partial_\tau g_{\lambda \mu }(x)\right]$,
in which  the matrix $g^{\mu\nu}(x)$ is defined as
$g^{\mu\nu}(x) = h^{i\mu}(x)h^{i\nu}(x)$.
Clearly the found relation is identical to the usual
expression for the Christoffel symbol
in terms of the Riemann metric for some Riemann geometry,
see for definitions Landau and Lifshitz.~\cite{LL}
Moreover, 
it is easy to verify that the quantity
$R^{\lambda}_{\rho\mu \nu}(x)$ can be presented in 
terms of $g_{\mu\nu}(x)$ as well
$
R_{\lambda\mu\nu}^{\lambda}(x)=
\partial_\mu \Gamma_{\lambda\nu}^{\lambda}-
\partial_\nu \Gamma_{\lambda \mu}^{\lambda}+
\Gamma_{\tau \mu}^{\sigma}\Gamma_{\lambda\nu}^{\tau}-
\Gamma_{\tau\nu}^{\sigma}\Gamma_{\lambda\mu}^{\tau}.
$
One recognizes in this 
relation the usual connection between the Riemann
tensor and the Riemann metric. 
We see that equation (\ref{dsda})
shows that the introduced in (\ref{ghh}) quantity
$g_{\mu\nu}(x)$ can be considered as 
a metric for some Riemann geometry 
with the Christoffel symbol $\Gamma^\lambda_{\mu\nu}(x)$
and the Riemann tensor $R^\lambda_{\rho\mu\nu}(x)$.

Consider now the second classical equation (\ref{dsdh}).
It is easy to verify that it can be presented
in the following form
\begin{equation} \label{einnom}
R_{\mu \nu} - \frac{1}{2} g_{\mu \nu} R = 0,
\end{equation}
in which
$R_{\mu\nu}(x)$ and $R(x)$ are the
Ricci tensor and the curvature of the Riemann geometry.
The found equation (\ref{einnom}) is identical to the
Einstein equations of general relativity in the absence of matter.

Moreover, the effective action (\ref{hhF})
may also be presented in geometrical terms.
To see this consider 
the action  when (\ref{dsda}) 
is valid. It is clear from (\ref{ghh}),(\ref{RF}) 
that the integrand of the action  (\ref{hhF}) proves be
proportional to the integrand of the usual
Hilbert action of general relativity \cite{LL}.
One can consider the  action 
(\ref{hhF}) and the Hilbert action  as same quantity
if the Newton gravitational constant $k$ is identified with
the constant $f$ which characterize the density of the 
polarized condensate of instantons
$k = 1/(16 \pi f)$.
This relation shows that $f=2/r_{\rm P}^2$. 
Remember that the constant 
$f$ introduced in (\ref{M^AB}),(\ref{M=fM})
depends on the typical
radiuses and separations of the polarized instantons.
We see that  these radiuses and separations should be
comparable with the Planck
radius.

\section{Discussion of results}
\label{res}

We come to the interesting result.
The first  classical equation of motion 
for the gauge field (\ref{dsda}) shows
 that particular combinations of the gauge field variables
(\ref{ghh}),(\ref{gam}),(\ref{RF}) 
are identical to the Riemann metric,
the Christoffel symbol, and the Riemann tensor for some Riemann space.
The second classical equation  (\ref{dsdh})
proves be identical to the Einstein equations for this Riemann metric.
Validity of the Einstein equations guarantees that
long-range excitations 
are massless spin-2 excitations.
The found effective action turns out to be identical to the  
Hilbert action of general relativity.
These facts altogether permit one to identify the found excitations 
with gravitons.
This indicates that
gravity arises in the framework of the gauge theory.
It is very important that  the dynamics of general relativity,
its action and equations of motion,
originate directly from the dynamics of the gauge field.
All these results follow  from 
assumption  (\ref{M=fM}) which 
has been interpreted above as the $SO(3,3)$ polarization of instantons.

In order to justify the considered scenario one needs to
find gauge models in which polarization of instantons
takes place. The necessary model
should satisfy several demanding conditions.
One of them is a necessity that
the polarization of (anti)instantons remains 
non-trivial even in the simplest case when gravitational field is absent.
In this case the polarization of instantons 
simplifies to be a constant $SO(3)\times SO(3)$ matrix.
To meet this requirement 
the least one needs  is  a $SU(2)$ gauge theory 
model in which instantons  are polarized, while
antiinstantons remain  non-polarized. 
A candidate for such a model has been  proposed in 
.~\cite{det}


\section*{Acknowledgments}
This work is supported by the Australian Research Council.
The hospitality of the stuff of the
Special Research Center for the Subatomic Structure of Matter 
at the University of Adelaide,
where part of this work has been carried out, is acknowledged.

\end{document}